# Accelerating pathways to leadership for underrepresented groups in STEM

*New strategies to broaden opportunities for STEM researchers and professionals from underrepresented groups to drive decision-making roles across sectors from philanthropy, industry, to academia, and policy*

*Authors:*

*2030STEM. 2030STEM Inc., NY, NY USA*

*Jennifer D. Adams, PhD, 2030STEM Salon Series Editor; University of Calgary, Calgary, Alberta, CAN*

*Cameron Bess, PhD, Project Officer and Biologist, U.S. Department of Health and Human Services; Board Member 2030STEM, Washington D.C., USA*

*Joshua C. Brumberg, Ph.D., Dean of the Sciences, The Graduate Center, CUNY, NY, NY, USA*

*Ruth Cohen, Interim Executive Director and Strategic Advisor, CoFounder 2030STEM, New York, New York, USA*

*Jacqueline K. Faherty\*, PhD, American Museum of Natural History; CoFounder 2030STEM, New York, New York, USA, ORCID 0000-0001-6251-0573*

*Daren R. Ginete, PhD, Associate Advisor, Science Philanthropy Alliance, Menlo Park, CA, USA*

*Mandë Holford\*, PhD, Hunter College; American Museum of Natural History; CoFounder 2030STEM, New York, New York, USA ORCID 0000-0001-9476-2687*

*Bobby Jefferson, Vice President and Global Head of Diversity, Equity, Engagement, and Inclusion, DAI, Washington D.C., USA*

*Jeanne Garbarino, PhD, Executive Director, RockEDU Science Outreach, The Rockefeller University, New York, NY, USA ORCID 0000-0002-8824-9294*

*Alfred Mays, Chief Diversity Officer and Strategist, Senior Program Officer – Diversity and Education Burroughs Wellcome Fund,*

*Chinyere Nwafor-Okoli, DVM, PhD, Provincial Epidemiologist, Alberta Trauma Services, Alberta Health Services, Foothills Medical Center, Calgary, Alberta.*

*Enrico Ramirez-Ruiz, PhD, Astronomy and Astrophysics Department, University of California Santa Cruz, Santa Cruz, California, USA*

*\*Corresponding authors: jfaherty@amnh.org ; mholford@hunter.cuny.edu*

**ABSTRACT**
The vision of 2030STEM is to address systemic barriers in institutional structures and funding mechanisms required to achieve full inclusion in Science, Technology, Engineering, and Mathematics (STEM) and accelerate leadership pathways for individuals from underrepresented populations across STEM sectors. 2030STEM takes a systems-level approach to create a community of practice that can test, learn and promote programs and policies that affirm and value cultural identities in STEM.

To achieve parity and full representation in the STEM workforce, a variety of changes are needed across academia and STEM professional industries (e.g., business, finance, biotech, government) to accelerate underrepresented groups into positions of leadership throughout the STEM ecosystem. Through a series of subject matter interviews, roundtables, and curated analysis four major themes have surfaced, which, if implemented, could exponentially accelerate the creation of critical pathways to leadership, break down pre-existing barriers and biases, intentionally elevate the voices, value, and research of underrepresented groups in STEM, and implement new structural strategies at scale. This white paper provides a summary of innovative new practices designed to accelerate inclusion, including expanding on known global toolkits, new funding strategies, and the structural changes required throughout various STEM professions to propel pathways to leadership for underrepresented groups.

This is the third in a series of white papers based on 2030STEM Salons that bring together innovative thinkers invested in creating a more equitable and inclusive STEM world for all. Our first white paper focused on the power of social media campaigns to accelerate change towards inclusion and leadership by underrepresented communities in STEM: [#Change: How Social Media is Accelerating STEM Inclusion](#). Our second white paper explores the various cultural shifts required across the mentoring ecosystem to support underrepresented groups in STEM along with an overview of several evidence-based mentorship techniques that provide a pathway to better overall experiences for the mentor and mentee: [Accelerating and Scaling Mentoring Strategies to Build Infrastructure that Supports Underrepresented Groups in STEM](#).

**OVERVIEW**

Decades of data continue to stack up showing the many values of a diversified leadership structure found in all industries and specifically those focused on STEM topics, yet Board Rooms, Dean's Offices and C-suite positions continue to not reflect the commitment to diversity outlined in many charters and bylaws. Frustrated by the many systematic barriers that have kept talented people out of the boardroom and out of senior leadership positions for generations, there is a renewed commitment to identify concrete pathways to leadership and access to funding for underrepresented groups in STEM. In a salon convened by 2030STEM entitled *"Accelerating pathways to leadership for underrepresented groups in STEM"*, effective strategies for designing and testing approaches that can rapidly improve retention and create pathways to leadership for underrepresented groups in STEM were discussed. Invited members of the salon, which included high-level decision-makers from academia, industry, and government agencies, analyzed the barriers and shared evidence-based opportunities for dismantling those barriers.

While there has been progress in education and training opportunities for more STEM researchers and professionals from underrepresented groups, significant change is still needed to accelerate the attainment of decision-making and leadership roles. As members of the STEM enterprise, our ultimate goal is to advance scientific findings, and diversity, equity, and inclusion (DEI) is an incredibly important component in enabling the successful achievement of that goal[1] [2]. In the 2030STEM Salon, several participants underscored the significance of DEI leadership in the scientific community throughout their discussions. For example, Dr. Enrico Ramirez-Ruiz outlined how diversity is inextricably linked to achieving scientific excellence; Christopher Perkins shared business models from the McKinsey Institute for Black Economic Mobility that can be replicated to accelerate change and is surfacing effective practices and learnings for recruiting and retaining Black talent in industry; Bobby Jefferson emphasized the importance of succession planning with an intentional focus on bringing diverse candidates into positions of leadership; and Dr. Daren R.

> "Many take progress for granted, and believe that issues of discrimination, bias, and inequality will subside naturally with time, yet progress towards an equitable feature is not inevitable – it requires intentionality, vigilance, and a commitment to concrete and sustained action." – Chen *et al,.* 2022 (10.31219/osf.io/xb57u)

Ginete discussed how emerging shifts in funding strategies will enable research and accelerate career pathways for underrepresented groups in STEM.

Leaders are essential to ushering in meaningful change across both the academy and in industry. With widespread support from faculty or industry leadership, the change needed to disrupt and dismantle existing structures and barriers to equal opportunity can be accelerated and realized. Four significant themes emerged from these discussions to accelerate pathways to leadership, which we will expand on in this white paper:

- Guiding principles are needed to illuminate the importance of driving DEI in STEM
- Replicating innovative practices from industry through the STEM ecosystem may accelerate change at scale
- An integrated strategy that leverages faculty, research and metrics is needed to drive structural changes in academia
- Funding strategies and outcomes should propel a shift to enabling research and career pathways for underrepresented groups in STEM

**GUIDING PRINCIPLES FOR DIVERSITY, EQUITY, AND INCLUSION IN STEM**
The pursuit of science and scientific excellence is inseparable from the people who populate it and lead it. As recently described, "scientific research is a social process that occurs over time with many minds contributing[1]". This leads us to the question, "How do we define scientific excellence?" If we think the answer is that scientific excellence is the equitable optimization of knowledge, infrastructure, and innovation, then this, of course, should include not only the technical aspects but also the non-technical contributions and views of all partners. We know from the Diversity-Innovation Paradox that heterogeneous groups produce higher quality and more innovative outcomes[3]. By this definition, true scientific excellence is not possible without equitable participation. We are never going to be truly scientifically excellent in any field if an amalgamation of voices are not at the table.

*Change the paradigm around the word "excellence":* When it comes to candidate selection and promotion, the word "excellence" can become a gatekeeping mechanism. The definition of excellence is often based on historic or culture-specific terms that do not account for the breadth of lived experiences of underrepresented staff. Organizations requiring applicants to have a full package to meet their definition of excellence run the risk of perpetuating barriers to progression. When this occurs, institutions and/or mentors differ in spending time and resources to upskill potential candidates to allow them to grow into leadership roles to excel in a new environment.

*Lean into mentoring as an accelerator:* If DEI is truly desired by an institution or organization, then deep mentoring and embracing faculty of diverse backgrounds should be centered, valued, and resourced appropriately. Recognize that some skills should be learned on the job and that candidates don't need to come in 100% ready. It's also important to focus on the strengths of these individuals, rather than their perceived gaps, to shift the mindset and identify how to best leverage the skill set they possess now. Leaders should also recognize that predominantly white faculty may require training for a new mentoring paradigm to ensure mentees from underrepresented groups receive effective support and value from the mentoring relationship. Even well-meaning mentors may not possess the skills needed to fully support members of communities of which they are not from[4] [5]. Mentor training would help systematically change leaders in the promotion of future leaders. Learn more about Accelerating and Scaling Mentoring Strategies to Build Infrastructure that Supports Underrepresented Groups in STEM.

*View "survival skills" as a value add:* Faculty and industry leaders should appreciate that soft skills some candidates from underrepresented groups have had to acquire to "survive" are additive and bring significant value to leadership positions. These skills often reflect grit, determination, and a drive to succeed that oftentimes comes with an expectation of assimilation to an environment that does not champion individuality and negates a true sense of belonging and inclusiveness.

**REPLICATING INNOVATIVE PRACTICES FROM INDUSTRY TO ACCELERATE CHANGE AT SCALE**
Unlike academia or government agencies, STEM industries can be more nimble, allowing their policies to pivot to create new cultures that drive innovation and profits. Industry's evolving policies and investments to include DEI efforts have allowed them to respond to often neglected communities in STEM. In doing so industry leaders have embraced key tenets:

*Investment requires more than money:* Investment should be more than just financial support. True change can only happen when people, time, and resources are invested in changing the process and the outcome. Investing in capacity strengthening and deepening ties between Minority Serving Institutions and communities with research-heavy STEM institutions can substantially bend the curve in promoting pathways to leadership roles.

*Track diversity in the workforce:* Track the importance of diversity in the workforce and uncover ways to demonstrate the impact of individuals from underrepresented groups in leadership. Below are several examples of innovative practices to foster retention and leadership roles for underrepresented groups in STEM.

<u>Lessons from the McKinsey Institute for Black Economic Mobility</u>
In response to the racial reckoning of 2020, McKinsey started a firm-wide initiative that outlined 10 actions toward racial equity. Part of that strategy included developing A Black Leadership Academy and an Institute for Black Economic Mobility to help bring awareness to various racial inequities, and more importantly, to start facilitating solutions. In its work, the Institute for Black

Economic Mobility tracks how funding devoted to racial equity is used, develops research and disseminates insights via reports. They also hold convenings such as the Black Economic Forum on Martha's Vineyard, and have launched the McKinsey Black Network to support and retain its own Black talent.

Similar to a recent National Academy of Sciences, Engineering, and Medicine report[6], a recent study from the Institute spotlighted the importance of increasing representation and tech leadership roles for Black talent[7], and how technology can influence opportunities for the Black community. McKinsey found that technology companies with Black leadership outperformed in economic profit measures against the Fortune 500 as a whole[8]. While the data shows very positive economic outcomes, there remains a pipeline problem caused by three main challenges: First, there remain roadblocks related to sourcing Black talent, blockage in the pipeline, and there is also leakage in the pipeline. Second, there is a mismatch in Black talent supply and where companies are searching for talent. Finally, the talent exists, it just is not in the places that companies look, likewise, companies have not invested in building capacity in these communities.

The McKinsey Institute also found that there are two major discrepancies between Black tech talent and tech roles. The first is that states like California and Texas, which lead the country in terms of the number of tech roles, lack significant representation of Black tech talent. Secondly, states like Mississippi, Louisiana, South Carolina, and Washington D.C. offer a large pool of Black tech talent, however, the supply of tech roles is insufficient[8]. When it comes to leakage, if there aren't adequate opportunities, Black talent is leaving. Increasingly, companies are taking responsibility to address these challenges by forming affinity groups and providing space for these groups to meet with leadership to voice their concerns. To address the leakage challenge, McKinsey has begun prompting discussions with its executives about how they process losing talent and if they attempted to adjust team culture and roles in an effort to retain employees from underrepresented groups. While some "old school types" shared that they did not feel it was a special problem if a promising employee from an underrepresented group leaves, for others it prompts a discussion of whether or not the employee left because the company's socio-cultural environment was not conducive to their ability to achieve adequate career growth. To counter the former, affinity groups within McKinsey take a proactive approach to understanding in more detail why an employee is seeking opportunity elsewhere and document those trends into data for the Board and management.

Partnerships between for-profits, non-profits, and the government can build skills and capacities along career trajectories, in addition to providing a mentorship ecosystem. For instance, Google's Grow with Google HBCU Career Readiness program was launched in 2020 to support Black jobseekers. The program partners with the Thurgood Marshall College Fund along with HBCU career services centers to help Black college students develop the technical skills needed to find and secure internships and jobs. Google has shared that 75% of program graduates report an improvement in their career within 6 months of certificate completion[9].

Within McKinsey, the organization is starting to look at intersectionality as it relates to DEI. Additionally, the company is formalizing and intentionally assigning mentors and sponsors, from day one when someone is recruited to the firm spanning the entirety of their career. Beyond the company-assigned mentor, Black employees also receive a mentor within the McKinsey Black Network who is responsible for their growth and development in the firm.

Models for Succession Planning

It is important to build a model of succession planning designed to intentionally promote diverse talent into leadership roles. During the succession planning process, an organization should identify a pool of "high-potential" staff members from diverse backgrounds to consider for succession roles. The process of identifying and expanding diverse talent is core to the organization's ongoing success of diverse leadership. The commitment to diversity happens at the board and senior leadership level where teams work to build the organization's leadership pipeline.

One hurdle organizations face is the use of the "high-potential" phrase, which sometimes becomes a barrier to opportunity for staff from underrepresented groups if they do not have an ally or mentor advocating for them as a high-performing, high-potential individual. "High potential" is a vague label that can narrow the pool of skilled candidates and opens up the possibility of bias toward individuals without this distinction. According to Brandon Hall Group's research study *Developing Your High-Potential Talent*, 60% of respondents said that their organization's high-potential identification and development processes are biased[10]. This is problematic because how organizations define potential affects who gets promotions, leadership coaching, and other development opportunities.

It is important to provide STEM leadership coaching and mentoring to next-generation STEM leaders. Access to coaching is particularly important for individuals across the organizations from underrepresented groups to help them build additional skills that will gain recognition when being considered for leadership positions in the future that are determined by the succession planning process.

**LEVERAGE FACULTY, RESEARCH, AND METRICS TO DRIVE STRUCTURAL CHANGES IN ACADEMIA**

It is widely recognized that structural change is needed across the academic ecosystem to dismantle systemic barriers that have historically prevented underrepresented groups in STEM from securing funding and publication opportunities, and ultimately, rising to leadership positions. In some sectors of STEM, the situation is incredibly dire regarding the pipeline of researchers from underrepresented groups. The American Astronomical Society's' survey in Astronomy and Astrophysics included a sub-committee tackling science in a social context which found incredibly low numbers of Black and Latinx researchers[11]. In the astronomy field, Black and Latinx researchers comprise just 3% of the faculty, while there are just two indigenous faculty members across the 38 departments surveyed. Just one astronomy department has both an African American and Latinx faculty member, while 21 departments have no diverse faculty members at all. A pathway to changing these demographics includes leveraging existing and new faculty, research, and metrics that amplify members in underrepresented STEM communities.

Partnering with Minority Serving Institutions (MSIs) can create STEM innovation through heterogeneity and importantly, mine the community-focused work that MSIs have been doing for years to surface and inform research goals for science in society. By intentionally elevating and amplifying the voices at MSIs, as a research community, the broader STEM ecosystem can improve science research and the scientific enterprise through DEI.

However, relying on MSIs alone without restructuring Predominantly White Institutions (PWIs), which are the majority in academia, would limit the reach and sustainability of DEI and innovation efforts. Where there are clear structural inequities creating barriers to full participation, these inequities should be removed. By increasing the weighted value of diverse experiences, organizations can open pathways and opportunities for all by revising hiring, promotion and leadership recruitment policies and programmatic frameworks to redefine what is a "high potential" or "excellent" candidate.

Beyond restructuring research and elevating the voices of academics from underrepresented groups, there is an opportunity to adjust various academic mechanisms to accelerate DEI in the STEM ecosystem:

*Design new metrics:* Design new metrics to support substantial collaboration, targeted capacity strengthening of underrepresented students' labs and departments, and substantial opportunity generation for progression to and sustaining in leadership positions. In the funder space, there is an opportunity to shift the metrics used to evaluate success when applying for and awarding grants. Models that measure research outputs beyond publications should be considered. Things like service on thesis committees which enhance training, or on editorial boards of STEM entities that serve the community, or organization of conferences to share knowledge also bring significant value and should be considered as a metric used to measure success.

*Incentivize faculty participation:* Change academic faculty workload to provide more time and emphasis on things like mentoring and training that will accelerate leadership and growth for the next generation of STEM candidates. Explore incentives that would encourage full participation from busy or high-level mentors (i.e. grant over-head relief; teaching or Chair relief; travel and training, etc.).

*Create a catalyst for wage increases:* Graduate students and postdoctoral scholars should be paid a living wage. This is critically important for the recruitment and retention of students from underrepresented groups who are more likely than their white peers to face additional financial hurdles. Funding agencies and foundations can be a catalyst as we have seen the living stipend rise after NSF increased its graduate and postdoctoral fellowship stipends forcing other organizations to remain competitive. Recently the Howard Hughes Medical Institute initiated a [pay raise for its postdoctoral scholars](), which will have a ripple effect throughout the biomedical research field.

*Providing more lifestyle support:* Non-traditional students are predominantly from underrepresented groups and they may need various types of lifestyle support, including on-site childcare, that isn't typically found across the academic ecosystem. Enrichment grants should also be considered to support students, postdoctoral scholars, and even early- or mid-career faculty from underrepresented groups who are experiencing some kind of financial crisis.

The Lamat Institute: a case study for culture change in academia

The Lamat Institute of University of California, Santa Cruz, dedicated to transforming STEM by providing opportunities for early career scientists to engage in novel research and creating healthy spaces for scientific inquiry, provides a useful model for effective leadership pathway programs. Lamat, meaning "star" in Mayan, supports students in community colleges and HBCUs by tackling inequality, removing barriers to ensure scientific excellence in astronomy, and helping students transfer to four-year institutions. The Lamat Institute's design marries strong technical support of students with an intentional examination of power and identity to create equity-minded mentoring and resources for students, empowering students to unsettle the current power structures. Additionally, the program emphasizes a community of care, creating healthy research spaces that are welcoming to different identities and cultures, amplifying all voices, and intentionally creating opportunities for growth and promotion. Importantly, the program has been designed to take a very pedagogical approach to introducing research, done effectively to help students thrive. This restructuring of introduction to research courses is critical as this shift reflects challenges faced by students from underrepresented groups.

In more than a decade, the Institute has served 110 students and made measurable progress in the number of graduates from historically underrepresented groups. For example, among Latinx students who make up a large percentage of California community colleges, Lamat has helped to double the number of Latinx students enrolling in top graduate schools across the country. Of the 110 Lamat Institute students, 85% have gone onto graduate school while 40% have published a first-author paper. The Lamat Institute provides clear evidence that initiatives can be both scientifically excellent and lead with care, creating spaces where science is done in a way that accepts and welcomes people of all backgrounds.

**FUNDING STRATEGIES AND OUTCOMES should ADAPT TO ACCELERATE STEM LEADERSHIP CHANGE**

New funding strategies and outcomes are emerging and should scale to propel sustained and prolonged investment that enables research and career pathways for underrepresented groups. It will take several generations of sustained effort to change the status quo, and funders will need to be committed for the long haul if they want to see real and exponential change. For instance, private philanthropy can give endowment funds that allow for the long-term and sustained investments required to bend the curve and accelerate change where federal funding cannot.

Funding strategies are beginning to intentionally shift to support research by scientists from underrepresented groups in more diverse numbers and types of universities and colleges, including HBCUs and MSIs. A continued partnership among private philanthropy and public funding can propel the shift, which enables research and career pathways for underrepresented groups. Despite this progress, a recent landscape report commissioned by the Alfred P. Sloan Foundation indicated that 44% of philanthropic investments to STEM higher education go to a mere 10 institutions, further perpetuating barriers to DEI. It is critical for the scientific community to break the cycle of poor funding and the ripple effect this causes downstream such as fewer publications, which in turn leads to fewer grants.

3 Categories of DEI Investments and Emerging Trends that Guide Funding
There are three main categories that DEI investments fall into, comprising pillars that guide the Science Philanthropy Alliance's DEI investments: (1) supporting programs of grantees such as identifying specific funding areas toward DEI. (2) changing processes with a goal to drive more equitable distribution of funds. (3) creating programs such as Howard Hughes Medical Institute's (HHMI) Freeman Hrabowski Scholars;The HHMI Freeman Hrabowski Scholars program supports outstanding early career faculty who are committed to advancing DEI in science. The fund provides legitimacy for the work and in a way that it is not additional labor required of faculty from underrepresented groups.

Three trends are emerging that will also guide funding interests:

1. *Focus on early and mid-career researchers to change the face of scientific leadership:* Philanthropy has a growing interest in supporting mid-career researchers as there is alignment between philanthropic investments and an opportunity to diversify scientific leadership. This trend empowers funders to have more decision-making capabilities.
2. *Partnerships with other entities:* Mutually beneficial partnerships are increasingly emerging between MSIs and predominantly white institutions (PWIs), along with private-public partnerships. MSIs are a valuable resource for increasing STEM inclusivity, however, they often experience funding gaps and need to tap into a variety of funding sources to secure adequate investment. For instance, the National Science Foundation's Chan Zuckerberg Initiative has partnered with the National Academies of Sciences, Engineering, and Medicine to launch the Science Diversity Leadership program. Together, they are providing a funding opportunity that aims to further the leadership of biomedical researchers who — through their outreach, mentoring, teaching, and leadership — have a record of promoting DEI in their scientific fields.
3. *Shift to participatory grantmaking:* Grants that require and include community member participation as part of the decision-making process on funding choices represent another frontier for science philanthropy to consider. While there is promise that much of DEI-related investments are going to organizations led by people of color, shifting decision-making to community members still lags. A report on participatory grantmaking is being developed by the Simons and the Heising-Simons Foundations.

**RECOMMENDATIONS TO ACCELERATE LEADERSHIP PATHWAYS AT SCALE**
To unlock and accelerate leadership pathways for underrepresented groups in STEM, we should properly support people of all backgrounds engaged in the STEM ecosystem. Not only will this unlock scientific innovations, but it will establish healthy cultures for inquiry and allow a full range of human diversity to meaningfully contribute to the field of science. To dismantle pre-existing barriers in institutions and organizations, there is a need for intentionality in ensuring practices match stated DEI values. It should invest a variety of resources, including people, time, and funding, to shift long-term outcomes. Investments across each phase of the pipeline are critical to creating meaningful, long-term impact

Industry Leadership Pathway Models to Replicate and Scale

- *Building the talent pipeline:* Building STEM skills should start early, requiring programs that support elementary, middle, and high school students. Companies will want to be tightly engaged in their local community, investing financial resources and time to begin building a pipeline of K-12 talent. Connections to local schools and being engaged in afterschool programs will bring students from underrepresented groups' level of achievement up and familiarize students with the companies and potential STEM careers
- *Integration with private sector affinity groups:* The private sector is heavily investing time and energy into employee affinity groups, and there is an opportunity to connect these employee/minority resource groups within private sector companies with students from underrepresented groups. Connections between affinity groups and students helps students realize opportunities at these private sector organizations in terms of internships, entry-level positions, or other opportunities for mentorship and networking. Fostering a connection between students from underrepresented groups to the range of different funding sources and research opportunities available from the private sector empowers institutions to pave leadership pathways for students.
- *Expand the Board:* Whether in academia or industry, there is an opportunity to add new voices by creating more board seats. Often long-standing board members are asked to cede their seat on the board to make room for a member from an underrepresented group, but this tactic creates resentment and strips a board of its institutional history. Instead, after preparing candidates from an underrepresented group to hold a board position, the senior members of the board should then guide the recently added members through the onboarding process to ensure the vision and mission of the board is duly communicated and retained.
- *Be intentional in succession planning:* Implement succession plans that have an equal number of candidates from underrepresented groups. It is a best practice to include a DEI officer in assessing leadership pathways for employees and identifying those with the greatest potential, and include training and mentorship to support the success of the diverse talent in this role.

Institutional Policies to Restructure Research and Mechanisms to Drive Culture Change

- *Prioritizing DEI work in promotion and tenure packages:* Often faculty from underrepresented groups are doing most of the DEI work, including mentoring not only their students but other students through shadow mentoring, along with volunteer work on DEI committees. However, when being evaluated for promotion and tenure, administration is more focused on the amount of papers published and grants awarded. As an academic community, we should find value in this extra DEI work and make it more of a priority.
- *Deploy partnerships and test programs to accelerate change:* Identify opportunities to scale leadership pathway programs that lead to diverse candidates achieving meaningful roles. There is an opportunity to deploy replications of The Lamat Institute program and test it in major metropolitan areas for five years to help early career scientists from underrepresented groups thrive in their studies.
- *Show commitment through budget:* Investment priorities speak to an institution's interests and values. If there is no capital investment in DEI, this questions how important DEI is for the institution. Allocating meaningful funds to DEI programs can create pathways for leadership roles from underrepresented groups.

Funding Guidelines To Distribute Access to Capital to Promote DEI Leadership Across Academia

- *Inclusive practices in grant design:* We should advocate for consistent, measurable, and research-proven practices in grant design to accelerate change. Requirements that add diversity "on paper" need to be reevaluated. True partnerships between Minority Serving Institutions and Predominantly White Institutions flow through dedicated funding, exchanges, authorship, and positions of leadership. Future diversity requirements should increase accountability on the prime recipient to show exponential change through partnerships that elevate underrepresented groups in STEM. Creating accountability by funders may involve such strategies as changing the definition of "excellence in science" to embrace effective mentoring and other inclusive skills and practices.
- *Increased funding for MSIs and underrepresented students:* Funding research for MSIs provides critical lab experiences and research that underrepresented students can speak to when applying for graduate programs or industry positions. Having access to this foothold into the pipeline is critical. The disparity in funding of PWIs has to be rethought to support underrepresented groups in STEM throughout an array of environments.
- *Investing in late-stage scientists from underrepresented groups:* When someone more established in their career leaves the STEM field, we lose them and their expertise, and we also remove them from the STEM community. Despite not having all of the funding they need, these researchers have advanced in their careers and are contributing valuable work in their fields. These late-stage researchers serve are beacons requiring support.

Imagine the possibilities for inclusive STEM if we as an academic community were able to identify, replicate, and scale leadership pathway programs in new settings.

REFERENCES

[1] Thorp, HH. (2023) It Matters Who Does Science. *Science Editors Blog May 11, 2023*
https://www.science.org/content/blog-post/it-matters-who-does-science

[2] Nature Editorial Team. *(*2023) Why *Nature* is Updating Its Advice to Authors on Reporting Race or Ethnicity. *Nature* 616, 219. doi: https://doi.org/10.1038/d41586-023-00973-7

https://www.nature.com/articles/d41586-023-00973-7

[3] Hofstra B, Kulkarni VV, Galvez SMN, He B, Jurafsky D, McFarland D. (2020) The Diversity-Innovation Paradox in Science. PNAS 117 (17) 9284-9291.

https://www.pnas.org/doi/10.1073/pnas.1915378117

[4] Martinez-Cola, M. (2020). Collectors, Nightlights, and Allies, Oh My. *Understanding and Dismantling Privilege 10(1), 61-82*. Retrieved from11

ACKNOWLEDGMENTS

2030STEM Inc. gratefully acknowledges funding from the Alfred P. Sloan Foundation (G-2021-16977) and for their inspirational support in our planning year and for our Salon series. 2030STEM also acknowledges all participants of the #Change Salon for their thoughtful insight, visionary contributions, and dedication to building a STEM ecosystem that works for all. Holford's work was also supported by National Science Foundation Award (DRL # 2048544).